\documentstyle[epsf,aps]{revtex}

\begin{document}

\preprint{DPNU-98-46}

\title{Dynamics in the Conformal Window in QCD like Theories}

\author{V.A.~Miransky}

\address{Bogolyubov Institute for Theoretical Physics,
252143, Kiev, Ukraine\\
and Department of Physics, Nagoya University,
Nagoya 464-8602, Japan \\}

\date{\today}
\draft

\maketitle

\begin{abstract}
The dynamics with an infrared stable fixed point 
in the conformal window in QCD like theories
(with a relatively large number of fermion flavors) is studied.
The dependence
of masses of colorless bound states on a bare 
fermion mass is described. In particular it is shown that
in such dynamics,
glueballs are much lighter than bound states composed
of fermions, if the value of the infrared fixed point is not too
large.
This yields a clear signature for the conformal
window, which in particular can be useful for lattice
computer simulations of these gauge theories.
\end{abstract}

Recently, there has been considerable interest in the existence of
a nontrivial conformal dynamics in 3+1 dimensional
non-supersymmetric vector like gauge theories, with
a relatively large number of fermion flavors $N_f$
\cite{1,2,3,4,5,6}.  
The roots of this problem go back to a work of Banks and Zaks \cite{7}
who were first to discuss the consequences of the existence of 
an infrared-stable fixed point $\alpha=\alpha^{*}$ for $N_f>N_f^{*}$ in 
vector-like gauge theories.  
The value $N_f^{*}$ depends on the gauge group:  
in the case of SU(3) gauge group, $N_f^{*}=8$ in the two-loop approximation.

A new insight in this problem \cite{1,2} has been, on the one hand, connected
 with using the results of the analysis of the Schwinger-Dyson (SD) equations
describing chiral symmetry breaking in QCD (for a review, see Refs.\cite{8,9})
and, on the other hand, with the discovery of the conformal window in $N=1$
supersymmetric QCD \cite{10}.

In particular, Appelquist, Terning, and  Wijewardhana \cite{1}
suggested that, 
in the case of the gauge group SU($N_c$), 
the critical value $N_f^{cr}\simeq4N_c$ 
separates a phase with no confinement and chiral symmetry breaking 
($N_f>N_f^{cr}$) and a phase with confinement and with chiral symmetry 
breaking ($N_f<N_f^{cr}$).  The basic point for this suggestion was 
the observation that at $N_f>N_f^{cr}$ the value of the infrared fixed 
point $\alpha^{*}$ is smaller than a critical value
$\alpha_{cr}\simeq\frac{2N_c}{N_c^2-1}\frac{\pi}{3}$,
presumably needed to generate the chiral condensate \cite{8,9}.

The authors of Ref.\cite{1} considered only the case when the running coupling
constant $\alpha(\mu)$ is less than the fixed point $\alpha^{*}$.  
In this case the dynamics is asymptotically free (at short distances)
both at $N_f<N_f^{cr}$ and 
$N_f^{cr}<N_f<N_f^{**} \equiv\frac{11N_c}{2}$.

Yamawaki and the author \cite{2} analysed the dynamics in the whole
($\alpha, N_f$) plane and suggested the ($\alpha, N_f$)-phase diagram of 
the SU($N_c$) theory (see Fig. 1 below).\footnote{This phase diagram
is essentially different from the original 
Banks-Zaks diagram \cite{7}. For details, see Sec.VII in Ref.\cite{2}}. 
In particular, it was pointed out that one can get an interesting 
non-asymptotically free dynamics when the bare coupling 
constant $\alpha^{(0)}$ is {\it larger} than $\alpha^{*}$, 
though not very large.

The dynamics with $\alpha^{(0)}>\alpha^*$ admits a continuum
limit and is interesting in itself.  
Also, its better understanding can be important for establishing 
the conformal window in lattice computer simulations of 
the SU($N_c$) theory with such large values of $N_f$.
In order to illustrate this, let us consider the following example.
For $N_c=3$ and $N_f=16$, the value of the infrared fixed point $\alpha^*$
is small:
$\alpha^*\simeq$0.04 (see below).  To reach the asymptotically free
phase, one needs to take the bare coupling $\alpha^{(0)}$ less than this
value of $\alpha^*$.  
However, because of large finite size effects, the lattice
computer simulations of the SU(3) theory with such a small $\alpha^{(0)}$
would be unreliable.  
Therefore, in this case, it is necessary to consider the dynamics with
$\alpha(\mu)>\alpha^*$.

One of the goals of this paper is establishing a clear signature of 
the existence of the infrared fixed point $\alpha^*$,
which would be useful for lattice computer simulations.
The signature we will suggest is the spectrum of low energy excitations
in the presence of a bare fermion mass.
In particular, we will show that in this case, unlike the familiar
QCD with a small $N_f$ ($N_f$=2 or 3), glueballs are much lighter 
than bound states composed of fermions, if the value of the
infrared fixed point is not too
large.
Another characteristic point is a strong (and simple) dependence of 
the masses of all the colorless bound states on the bare fermion mass,
even if the latter is tiny.

We begin by recalling the basic facts concerning the two-loop $\beta$
function in an SU($N_c$) theory.
The $\beta$ function is
\begin{equation}
\beta = -b\alpha^2 - c\alpha^3 
\label{beta}
\end{equation}
with \cite{11}
\begin{mathletters}
\begin{eqnarray}
b&=&\frac{1}{6\pi} (11N_c - 2N_f),
\label{b} \\
c&=&\frac{1}{24\pi^2} (34N_c^2 - 10N_cN_f - 3\frac{N_c^2 -
1}{N_c}N_f).
\label{c}
\end{eqnarray}
\end{mathletters}
While these two coefficients are invariant under change of 
a renormalization scheme, the higher-order coefficients are 
scheme dependent.
Actually, there is a renormalization scheme in which the two-loop 
$\beta$ function is (perturbatively) exact \cite{12}.
We will use such a renormalization scheme.

If $b>0$ ($N_f < N_f^{**} \equiv \frac{11N_c}{2}$)
and $c<0$, the $\beta$ function has a zero, corresponding to
a infrared-stable fixed point, at
\begin{equation}
\alpha = \alpha^* = - \frac{b}{c}.
\label{alpha^*}
\end{equation}

When $N_f$ is close to $N_f^{**}$, the value of $\alpha^*$ is small.
For example, from Eqs.(\ref{b}), (\ref{c}), and (\ref{alpha^*}), one
gets $\alpha^* \simeq$ 0.04,
0.14, 0.28, and 0.47 for $N_c$=3 and $N_f$=16, 15, 14, and 13, respectively.

The value of $\alpha^*$ becomes equal to
$\alpha_{cr} = \frac{2N_c}{N_c^2-1}\frac{\pi}{3}$
at $N_f$ close to $N_f\simeq4N_c$.
And the fixed point disappears at the value $N_f=N_f^*$, 
when the coefficient $c$ becomes positive 
($N_f^*$ is $N_f^*\simeq8.05$ for $N_c$=3).

The $\beta$ function (\ref{beta}) leads to the following solution for 
the running coupling:
\begin{equation}
b\log\left(\frac{q}{\mu}\right) = \frac{1}{\alpha(q)}
 - \frac{1}{\alpha(\mu)} -
\frac{1}{\alpha^*}\log\left(\frac{\alpha(q)(\alpha(\mu) - \alpha^*)}
{\alpha(\mu)(\alpha(q) - \alpha^*)}\right).
\label{solution}
\end{equation}
We emphasize that this solution is valid both for
$\alpha(\mu)<\alpha^*$ and $\alpha(\mu)>\alpha^*$.

Let us first consider the case with $\alpha(\mu)<\alpha^*$.
It is convenient to introduce the parameter [1]
\begin{equation}
\Lambda\equiv\mu\exp\left[-\frac{1}{b\alpha^*}\log\left(\frac{\alpha^* -
\alpha(\mu)}{\alpha(\mu)}\right) - \frac{1}{b\alpha(\mu)}\right].
\label{lambda}
\end{equation}
Then, Eqs. (\ref{solution}) and (\ref{lambda}) imply that
\begin{equation}
\frac{1}{\alpha(q)} = b\log\left(\frac{q}{\Lambda}\right)
 + \frac{1}{\alpha^*}
\log\left(\frac{\alpha(q)}{\alpha^* - \alpha(q)}\right).
\label{solution1}
\end{equation} 
Taking $q=\Lambda$, we find that
\begin{equation}
\frac{\alpha^*}{1 + e^{-1}}\simeq 0. 73 \alpha^*< \alpha(\Lambda) 
< \alpha^*.
\label{alpha(Lambda)}
\end{equation}
One may think that $\Lambda$ plays here the
same role as $\Lambda_{QCD}$
in the confinement phase.  
However, as we will see, its physical meaning is somewhat different.

Eq. (\ref{solution1}) implies that
\begin{equation}
\alpha(q)\simeq \frac{1}{b\log\frac{q}{\Lambda}}
\label{uv}
\end{equation}
for $q>>\Lambda$ (the usual behavior in asymptotically free
theories), and
\begin{equation}
\alpha(q)\simeq \frac{\alpha^*}{1 +
e^{-1}(\frac{q}{\Lambda})^{b\alpha^*}}
\label{ir}
\end{equation} 
for $q<<\Lambda$, governed by the infrared fixed point $\alpha^*$.

Let us turn to a less familiar case with $\alpha(\mu)>\alpha^*$.
One still can use Eq.(\ref{solution}).
Introduce now the parameter $\tilde{\Lambda}$ as
\begin{equation}
\tilde{\Lambda}\equiv \mu\exp\left[-\frac{1}{b\alpha^*}\log\left(
\frac{\alpha(\mu) - \alpha^*}{\alpha(q)}\right)
- \frac{1}{b\alpha(\mu)}\right]
\label{tilde}
\end{equation}
(compare with Eq.(\ref{lambda})).  Then, Eqs.(\ref{solution})
and (\ref{tilde}) imply
\begin{equation}
\frac{1}{\alpha(q)} = b\log\frac{q}{\tilde{\Lambda}} +
\frac{1}{\alpha^*}\log\left(\frac{\alpha(q)}{\alpha(q) - \alpha^*}
\right). 
\label{solution2}
\end{equation}
What is the meaning of $\tilde{\Lambda}$?  
It is a Landau pole at which $\alpha(q)\vert_{q=\tilde{\Lambda}}=\infty$.
Indeed, taking $q=\tilde{\Lambda}$ in Eq.(\ref{solution2}), one gets
\begin{equation}
\frac{1}{\alpha(\tilde{\Lambda})} =
\frac{1}{\alpha^*}\log{\frac{\alpha(\tilde{\Lambda})}
{\alpha(\tilde{\Lambda}) - \alpha^*}}.
\label{pole}
\end{equation}
The only solution of this equation is $\alpha(\tilde{\Lambda})=\infty$.

The presence of the Landau pole implies that the dynamics is not
asymptotically free.
To get a more insight in this dynamics, let us introduce an ultraviolet 
cutoff $M$ with the bare coupling constant 
$\alpha^{(0)}\equiv\alpha(q)\vert_{q=M}$.
Now all momenta $q$ satisfy $q\leq M$.

Eq.(\ref{solution2}) implies that at finite $\alpha^{(0)}=\alpha(M)$,  
the cutoff $M$ is less than $\tilde{\Lambda}$, with
$\alpha(\tilde{\Lambda})=\infty$.  
Therefore the Landau pole is unreachable in the theory with 
cutoff $M$ and with $\alpha^{(0)}<\infty$.
Still one can of course use $\tilde{\Lambda}$ (\ref{tilde}) for 
a convenient parametrization of the running coupling 
$\alpha(q)$ (see Eq.(\ref{solution2})). However, one should remember that
momenta $q$ satisfy $q\leq{M}<\tilde{\Lambda}$.

Eq.(\ref{solution2}) implies that
\begin{equation}
\alpha^2(q)\simeq \frac{\alpha^*}{2b\log\frac{\tilde{\Lambda}}{q}}
\label{uv1}
\end{equation}
for $\alpha(q)>>\alpha^*$, and
\begin{equation}
\alpha(q)\simeq \frac{\alpha^*}{1
- e^{-1}(\frac{q}{\tilde{\Lambda}})^{b\alpha^*}}
\label{ir1}
\end{equation}
when $\alpha(q)$ is close to $\alpha^*$, i.e. when 
$\alpha(q)-\alpha^*<<\alpha^*$.
Thus, now $\alpha(q)$ approaches the fixed poin $\alpha^*$
from above (compare with Eq.(\ref{ir})).
And, in general, Eq.(11) implies that $\alpha(q)$ 
monotonically decrease with $q$, from 
$\alpha(q)=\alpha^{(0)}$ at $q=M$ to $\alpha(q)=\alpha^*$ at $q=0$.

Does a meaningfull continuum limit exist in this case?
The answer is of course "yes". As it follows from
Eq.(\ref{solution2}), when $M$
(and therefore $\tilde{\Lambda}$)
gose to infinity, and the bare coupling $\alpha^{(0)}>\alpha^*$ 
is arbitrary but fixed, $\alpha(q)$ is equal to the fixed value, 
$\alpha(q)=\alpha^*$, for all $q<\infty$.
Therefore it is a non-trivial conformal field theory.

Sor far we considered the solution for $\alpha(q)$ connected
with the perturbative (and perturbatively exact in the 't Hooft
renormalization scheme \cite{12}) $\beta$ function (\ref{beta}).  However,
unlike ultraviolet stable fixed points, defining dynamics at
high momenta, infrared-stable fixed points (defining dynamics at
low momenta) are very sensitive to nonperturbative dynamics 
leading to the generation of particle masses.
For example, if fermions acquire a dynamical mass, they decouple 
from the infrared dynamics, and therefore the perturbative infrared 
fixed point (\ref{alpha^*}) will disappear.

The phase diagram in the ($\alpha^{(0)}, N_f$)-plane in this theory
was suggested in Ref.\cite{2}.  It is shown in Fig. 1.
The left-hand portion of the curve in this figure coincides with 
the line of the infrared-stable fixed points $\alpha^*(N_f)$
in Eq.(\ref{alpha^*}). 
It separates two symmetric phases, $S_1$ and $S_2$,
with $\alpha^{(0)}<\alpha^*$ and $\alpha^{(0)}>\alpha~*$, 
respectively.  Its lower end is $N_f=N_f^{cr}$ (with 
$N_f^{cr}\simeq 4N_c$ if $\alpha_{cr}\simeq\frac{2N_c}{N_c^2-1}\frac{\pi}{3}$):
at $N_f^*<N_f<N_f^{cr}$ the infrared fixed point is washed out
by generating a dynamical fermion mass.

The horizontal, $N_f=N_f^{cr}$, line describes a phase transition
between the symmetric phase $S_1$ and the phase with confinement
and chiral symmetry breaking.
As it was suggested in Refs.\cite{1,2}, based on a similarity of
this phase transition with that in quenched $QED_4$ \cite{8,9,13}
and in $QED_3$ \cite{14}, there is the following scaling law for $m^2_{dyn}$:
\begin{equation}
m^2_{dyn}\sim \Lambda^2_{cr}\exp\left(-\frac{C}
{\sqrt{\frac{\alpha^*(N_f)}{\alpha_{cr}} - 1}}\right)
\label{mdyn}
\end{equation}
where the constant $C$ is of order one and $\Lambda_{cr}$ is a scale
at which the running coupling is of order $\alpha_{cr}$.

It is a continuous phase transition with an essential singularity
at $N_f=N_f^{cr}$.  The characteristic point of this phase 
transition is that the critical line $N_f=N_f^{cr}$ separates
phases with essentially different spectra of low energy excitations 
\cite{1,2} and the different structure of the equation for the divergence of 
the dilatation current (i.e. with essentially different realizations
of the conformal symmetry) \cite{2}.
It was called the conformal phase transition in Ref.\cite{2}.

At present it is still unclear whether the phase transition 
on the line $N_f=N_f^{cr}$ is indeed a continuous phase transition 
with an essential singularity or it is a first order phase transition
\cite{3,6}.
However, anyway, the two properties (the abrupt change of the spectrum
of excitations and the different structure of the equation for the
divergence of the dilatation current in those two phases) have
to take place.

\begin{figure}[htbp]
\begin{center}
\epsfxsize=8cm
\ \epsfbox{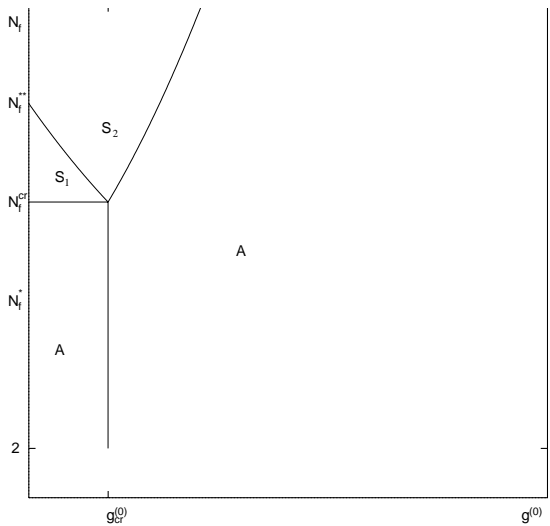}
\end{center}
\caption[]{The phase diagram in an SU($N_c$) gauge model. The
coupling constant $g^{(0)}=\sqrt{4\pi\alpha^{(0)}}$ and $S$ and
$A$ denote symmetric and asymmetric phases, respectively.}
\end{figure}

At last, the right-hand portion of the curve on the diagram occurs
because at large enough values of the bare coupling, spontaneous
chiral symmetry breaking takes place for any number $N_f$ of
fermion flavors.  This portion describes a phase transition called
a bulk phase transition in the literature, and it is presumably
a first order phase transiiton.
\footnote{The fact that spontaneous chiral symmetry breaking takes
          place for any number of fermion flavors, if $\alpha^{(0)}$
          is large enough, is valid at least for lattice theories
          with Kogut-Susskind fermions.
          Notice however that since the bulk phase transition is a 
          lattice artifact, the form of this portion of the curve
          can depend on the type of fermions used in simulations 
          (for details, see Ref.\cite{2}).}
The vertical line ends above $N_f$=0 since in pure gluodynamics 
there is apparently no phase transition between weak-coupling and
strong-coupling phases.

Up to now we have considered the case of a chiral invariant action.
But how will the dynamics change if a bare fermion mass term is
added in the action?
This question is in particular relevant for lattice computer
simulations:  for studying a chiral phase transition on a finite
lattice, it is necessary to introduce a bare fermion mass.
We will show that adding even an arbitrary small bare fermion
mass results in a dramatic changing the dynamics both in the
$S_1$ and $S_2$ phases.

Recall that in the case of confinement SU($N_c$) theories, 
with a small, $N_f<N_f^{cr}$, number of fermion falvors, 
the role of a bare fermion mass $m^{(0)}$ is minor if 
$m^{(0)}<<\Lambda_{QCD}$ (where $\Lambda_{QCD}$ is a confinement
scale).  The only relevant consequence is that massless 
Nambu-Goldstone pseudoscalars get a small mass (the PCAC dynamics).

The reason for that is the fact that the scale $\Lambda_{QCD}$,
connected with a scale anomaly, decribes the breakdown of the
conformal symmetry connected {\it both} with perturbative
and nonperturbative dynamics:  the running coupling and the
formation of bound state.
Certainly, a small bare mass $m^{(0)}<<\Lambda_{QCD}$ is 
irrelevant for the dynamics of those bound states.

Now let us turn to the phase $S_1$ and $S_2$, with $N_f>N_f^{cr}$.
At finite $\Lambda$ in $S_1$ and $\tilde{\Lambda}$ in $S_2$, there
is still conformal anomaly:  because of the running of 
the effective coupling constant, the conformal symmetry is broken.
It is restored only if $\Lambda\rightarrow{0}$ in $S_1$ and 
$\tilde{\Lambda}>M\rightarrow\infty$ in $S_2$.
However, the essential difference with respect to confinement
theories is that both $\Lambda$ and $\tilde{\Lambda}$ 
have nothing with the dynamics forming bound states:  
since at $N_f>N_f^{cr}$ the effective coupling is relatively weak, 
it is impossible to form bound states from $\it{massless}$
fermions and gluons (recall that the $S_1$ and $S_2$ phases are chiral 
invariant).

Therefore the absence of a mass for fermions and gluons is a key
point for {\it not} creating bound states in those phases.
The situation changes dramaticaly if a bare fermion mass is
introduced:  indeed, even weak gauge, Coulomb-like, interactions
can easily produce bound states composed of massive constituents,
as it happens, for example, in QED, where electron-positron
(positronium) bound states are present.

To be concrete, let us first consider the case when all fermions 
have the same bare mass $m^{(0)}$.
It leads to a mass function $m(q^2)\equiv{B(q^2)/A(q^2)}$ in the fermion 
propagator $G(q)=(\hat{q}A(q^2)-B(q^2))^{-1}$.
The current fermion mass $m$ is given by the relation
\begin{equation}
m(q^2)\vert_{q^2=m^2}=m.
\label{m}
\end{equation}

For the clearest exposition, let us consider a particular theory 
with a finite cutoff $M$ and the bare coupling constant
$\alpha^{(0)}=\alpha(q)\vert_{q=M}$
being not far away from the fixed point $\alpha^*$.
Then, the mass function is changing in the "walking" regime \cite{15} with 
$\alpha(q^2)\simeq\alpha^*$.  It is 
\begin{equation}
m(q^2)\simeq m^{(0)}\left(\frac{M}{q}\right)^{\gamma_m}
\label{m(q)}
\end{equation} 
where the anomalous dimension
$\gamma_m\simeq1-(1-\frac{\alpha^*}{\alpha_{cr}})^{1/2}$ \cite{8,9}.
Eqs.(\ref{m}) and (\ref{m(q)}) imply that
\begin{equation}
m\simeq m^{(0)}\left(\frac{M}{m^{(0)}}\right)^{\frac{\gamma_m}
{1 + \gamma_m}}.
\label{m1}
\end{equation}

There are two main consequences of the presence of the bare mass:

(a) bound states, composed of fermions, occur in the spectrum
of the theory.  The mass of a n-body bound state is 
$M^{(n)}\simeq{nm}$;

(b) At momenta $q< m$, fermions and their bound states decouple.
There is a pure SU($N_c$) Yang-Mills theory with confinement.
Its spectrum contains glueballs.

To estimate glueball masses, notice that at momenta $q< m$, the
running of the coupling is defined by the parameter $\bar{b}$ 
of the Yang-Mills theory,
\begin{equation}
\bar{b}= \frac{11}{6\pi}N_c.
\label{barb}
\end{equation}
Therefore the glueball masses $M_{gl}$ are of order
\begin{equation}
\Lambda_{YM}\simeq m\exp(-\frac{1}{\bar{b}\alpha^*}).
\label{YM}
\end{equation}

For $N_c=3$, we find from Eqs.(\ref{b}), (\ref{c}), and 
(\ref{barb}) that 
$\exp(-\frac{1}{\bar{b}\alpha^*})$
is $6\times{10^{-7}}$, $2\times{10^{-2}}$, $10^{-1}$,
and $3\times{10^{-1}}$ for
$N_f$=16, 15, 14, and 13, respectively.
Therefore at $N_f$=16, 15 and 14, the glueball masses are 
essentialy lighter than the masses of the bound states composed of
fermions.  
The situation is similar to that in confinement QCD with heavy quarks,
$m>>\Lambda_{QCD}$.
However, there is now a new important point:
in the conformal window,
{\it any} value of $m^{(0)}$ (and therefore $m$) is "heavy":
the fermion mass $m$ sets a new scale in the theory, and the 
confinement scale $\Lambda_{YM}$ (\ref{YM}) is less, and rather often 
much less, than this scale $m$.

This leads to a spectacular "experimental" signature of the 
conformal window in lattice computer simulations:
glueball masses rapidly, as 
$(m^{(0)})^{\frac{1}{1+\gamma_m}}$, 
decrease with the bare fermion mass $m^{(0)}$ for {\it all}
values of $m^{(0)}$ less than cutoff $M$.

Few comments are in order:

(1) The phases $S_1$ and $S_2$ have essentially the same 
long distance dynamics.
They are distinguished only by their dynamics at short distances:
while the dynamics of the phase $S_1$ is asymptotically free,
that of the phase $S_2$ is not.
In particular, when all fermions are massive (with the current mass $m$),
the continuum limit $M\rightarrow\infty$ of the $S_2$-theory is a 
non-asymptotically free confinement theory.
Its spectrum includes colorless bound states composed of fermions
and gluons.
For $q<m$ the running coupling $\alpha(q)$ is the same as in pure 
SU($N_c$) Yang-Mills theory, and for all $q> m$
$\alpha(q)$ is very close to 
$\alpha^*$ ("walking", actually, "standing" dynamics).
For those values $N_f$ for which $\alpha^*$ is small 
(as $N_f$=16, 15 and 14 at $N_c$=3), glueballs are much lighter than 
the bound states composed of fermions.
Notice that, unlike the case with $m=0$, there exists an S-matrix
in this theory.

(2) In order to get the clearest exposition, we assumed such estimates as
$N_f^{cr}\simeq 4N_c$ for $N_f^{cr}$ and 
$\gamma_m=1-{\sqrt{1-\frac{\alpha^*}{\alpha_{cr}}}}$ 
for the anomalous dimension $\gamma_m$. While the latter 
should be reasonalbe for $\alpha^*<\alpha_{cr}$ (and especially for
$\alpha^*<<\alpha_{cr}$) \cite{8,9}, the former is based on the
assumption that $\alpha_{cr}\simeq\frac{2N_c}{N_c^2 - 1}\frac{\pi}{3}$
which, though seems reasonable, might be crude for some values
of $N_c$.  It is clear however that the dynamical picture presented
in this paper is essentially independent of those assumptions.

(3) So far we have considered the case when all fermions
have the same bare
mass $m^{(0)}$.
The generalization to the case when different
fermions may have different bare masses is evident.

(4) Lattice computer simulations of the SU(3) theory with
a relatively large number of $N_f$ \cite{16,17} indeed indicate
on the existence of a symmetric phase.
However, the value of the critical number $N_f^{cr}$ is 
different in different simulations:
it varies from $N_f^{cr}=6$ \cite{17} through $N_f^{cr}=12$ \cite{16}.

We hope that the signature of the conformal window suggested 
in this paper can be useful to settle this important issue.

\begin{acknowledgments}
I thank M. Creutz, D. Sinclair, and K. Yamawaki
for useful discussions.
I wish to acknowledge the JSPS (Japan Society for the 
Promotion of Science) for its support during my stay at 
Nagoya University.
My special thanks to Koichi Yamawaki for his warm hospitality.
\end{acknowledgments}

\end{document}